\documentclass[11pt]{article}

\usepackage{epsfig}
\usepackage{graphicx}
\usepackage{rotating}
\usepackage{amsmath}
\usepackage{bm}
\usepackage[letterpaper, left=2.5cm, right=2.5cm, top=2.5cm, bottom=2.5cm]{geometry}

\usepackage{dirtytalk}
\usepackage{threeparttable}
\usepackage{tabularx}
\usepackage{booktabs}
\usepackage{placeins}
\usepackage{pdfpages}
\usepackage{multirow}
\usepackage{longtable}
\usepackage{url}
\usepackage{dcolumn}
\usepackage[style=apa]{biblatex}
\usepackage{xr-hyper}  
\title{Surprisingly-early bias in forecasts for unscheduled events%
  \thanks{I thank everyone who helped review this article, as well as Nikos Bosse and the Metaculus Team for data and data transformation support. No funds, grants, or other support were received for this work.}
}
\author{Niklas V. Lehmann%
  \thanks{Technical University Bergakademie Freiberg, Researcher at the Chair for General Economics and Macroeconomics. Contact: Niklas-Valentin.Lehmann@vwl.tu-freiberg.de}
}
\date{\today}

\begin{document}
\maketitle

\begin{abstract}
When a dataset contains forecasts on unscheduled events, such as natural catastrophes, outcomes may be censored or ``hidden'' since some events have not yet occurred. This article finds that this can lead to a selection bias which affects the perceived accuracy and calibration of forecasts. This selection bias can be eliminated by excluding forecasts on outcomes which have been verified surprisingly early.
\end{abstract}

\vspace{0.7em}

\vspace{0.2em}


\pagestyle{plain}
\setcounter{page}{1}


\section{Introduction}


When predicting the future, we seek to evaluate whether our forecasts are accurate and unbiased. A common way of evaluating whether forecasts are biased---i.e. whether they are systematically \say{off the mark}---is to look at the empirical reliability, often called \textit{calibration}. Calibration describes how well forecasts match real-world outcomes. That is, ``a forecaster is well calibrated if, for example, of those events to which he assigns a probability 30 percent, the long-run proportion that actually occurs turns out to be 30 percent'' \parencite{dawid_well-calibrated_1982}.

Yet, this seemingly straightforward standard falls apart when we turn our attention to events that are unscheduled or---more precisely---\textit{time-varying}. Unlike scheduled, or \textit{time-fixed} events, such as elections, weather on a certain day, GDP, energy demand, or opening weekend box office results, which are verifiable at a predetermined, single point in time, time-varying events can occur at more than one point in time. Natural disasters, pandemic outbreaks, technological milestones, deaths, or a team’s elimination from the playoffs are all examples of time-varying events.

The fundamental point of this article is that our position as observers of the past makes us encounter disproportionately more time-varying events that occurred surprisingly early than time-varying events whose timing was unexpectedly delayed, which introduces a selection bias.\footnote{We remark that \say{surprising} throughout this text is meant in an information-theoretical sense. Take, for instance, an event that is expected to occur \say{early} with a 99.9\% probability, and indeed does so.  Although not \say{surprising} in the everyday meaning of the word, it still---technically---counts as surprisingly early, as the Shannon entropy or surprisal is non-zero.}  For illustrative purposes, let's imagine a dataset of forecasts on a very large number of events that are each extremely unlikely to occur in the short-term. Some of them (if the number is large enough) will still occur in the short-term, and a sincere and accurate forecast was to judge the outcome as unlikely. Thus, the forecasts on these outcomes will---if observed in isolation---seem to have underestimated the probability of occurrence, even though the forecasts themselves may be perfectly accurate.

For example, in 2008 the U.S. Geological Survey published 30-year ahead earthquake forecasts for California \parencite{field_forecasting_2008}. If we evaluate the forecasts using data for earthquakes that have happened since 2008, we find that the study seems to have underestimated the actual frequency of seismic events.\footnote{The earthquake data can be found here: \texttt{https://earthquake.usgs.gov/earthquakes/search/}} For example, the Ridgecrest earthquake sequence in 2019 occurred with magnitudes up to 7.1 in an area where the 30-year probability of greater-than-6.7-magnitude earthquakes was predicted to be below 10\%. However, when we evaluate this forecasting data we only focus on the earthquakes that did happen, not those that did not happen. However, we cannot simply treat the unobserved earthquakes as not happened, since the forecasts describe cumulative risk over a 30-year horizon; additional earthquakes may still happen. There does not seem to be an alternative to waiting until 2038 to evaluate the accuracy of these earthquake forecasts.

In this study, we formalize what we term the \textit{surprisingly-early bias} in forecasting datasets. Through a straightforward simulation, we demonstrate that failing to exclude forecasts whose scheduled resolution dates extend beyond the data collection point results in selection bias. For instance, in the U.S. Geological Survey all forecasts are scheduled to resolve in 2038; attempting to analyze them prior to this date unavoidably introduces the surprisingly-early bias.
When we limit---or \say{filter}---our forecast dataset to include only those forecasts whose scheduled resolution dates precede the data collection point, the surprisingly-early bias is eliminated. 
To demonstrate the practical relevance of this effect, we analyze forecasting data from Metaculus, a reputation-based massive online prediction-solicitation platform. We find that the calibration of forecasts---as measured using all data---is distorted by a surprisingly-early bias, which makes clear that filtering is necessary to assess the calibration and accuracy of forecasts.

This is relevant when forecast data contains time-varying events, which unfold over time. Researchers, analysts, and practitioners working in domains such as natural disaster forecasting, pandemic forecasting, geopolitical forecasting, engineering and reliability analysis, and the actuarial sciences will potentially encounter the surprisingly early bias in their practice.

A similar problem to the one presented in this article can be found in survival analysis; sometimes also known as reliability analysis, duration analysis or event history analysis. This field, which is predominantly concerned with the timing of events, also frequently involves events that may not be fully observed \parencite{allison_event_2014}. For example, when forecasting the lifetimes of humans, we may obtain a dataset that contains humans that are still alive at the point of data collection. Thus, this \textit{censored} data---how long these humans will live---cannot be used in the analysis, yet simply excluding it can lead to selection bias \parencite{tuma_approaches_1979}. However, the problems studied in survival analysis are not quite analogous to the one that this article concerns. The problem in survival analysis usually concerns a \textit{population}, be it humans, machine parts, or even flowers. In this article we investigate forecasting data, which can refer to completely different and independent events; there is no population. Furthermore, censoring can have multiple causes in survival analysis, such as a participant failing to attend a follow-up survey, and can take various different forms. The censoring that we investigate here is far simpler: Forecasts are censored if the outcomes have not yet been verified. 

Rather, this article studies a form of survivorship bias, which is commonly known in many fields \parencite{brown_survivorship_1992}. A textbook example for survivorship bias is the study of damaged aircraft: we only ever observe damages that still allow a plane to return in a condition where it can be studied, thus inducing a systematic selection bias. Similarly, in our context, we only observe events that have already occurred, not those that linger in happening.

\textcite{hoga_time_nodate} discuss time-varying events in more in-depth and find that they pose different incentives to forecasters than time-fixed events. The reason for this is that time-varying events can happen surprisingly early, in which case forecasts who assign a higher probability to early potential outcomes are scored more highly earlier. If forecasters are present-biased, they will be compelled to strategically misreport their expectations to increase the chance of higher scores in the near-term future at the cost of lower scores in the long-term future. 
We stress that time preference effects can lead to real biases, i.e. forecasts change \textit{ex ante}, and that this bias cannot be easily eliminated. Conversely, the surprisingly-early bias is an artifact of data construction, only comes up \textit{ex post}, and can be eliminated easily.


This article proceeds as follows: 
Section 2 presents a simulation that clarifies the problem. Section 3 looks at the Metaculus forecasting platform, where we observe the surprisingly-early bias in real-world data. Section 4 discusses the results and concludes.

\section{Simulation study}\label{sec:Sim}

Let there be $N$ forecasts pertaining to $i \in [1,N]$ fully independent events, which can occur at any point in time. We denote a point in time by $t \in [0,T]$ and assume that all forecasts are made in $t=0$.  We denote a positive outcome---which can occur at any time---of the event $i$ by $X_i=1$. If the event did not occur by the scheduled resolution date $t_i \in [0,T]$ then $X_i=0$. In all other cases, $X_i$ remains unobserved. We assume that the forecasts take the simplest possible form, a binary probabilistic forecast:\footnote{We can generalize to all other forecasts---such as distributions---from the binary case, so the model is without loss of generality \parencite{gneiting_strictly_2007}. However, using distributional forecasts across time solves our problem. We defer to the discussion.} 

\begin{quote}
    What is the probability that the event $i$ will occur by $t_i$?
\end{quote}

We implement a hypothetical perfect forecaster; this forecaster knows the precise probability with which each event occurs at any point in time, i.e. the forecaster knows the density distribution function of the event.\footnote{
Since any event is still assumed to be random, the forecaster is not a perfect clairvoyant. A way to think of the perfect forecaster is to assume that they need to predict the outcome of a biased coin flip. A perfect forecaster is not a clairvoyant who can know what the outcome of the next flip is going to be with certainty. Rather, the perfect forecaster knows the bias exactly, and will be able to predict the frequency of coin flips with perfect accuracy as the number of coin flips approaches infinity. } 
 We sample $N=10^5$ random events whose occurrence shall be normally distributed across $t$.\footnote{The choice of distribution is quite arbitrary for the argument presented here, as long as it does not have a point mass of 1 at some point, that is, the event is truly random.} The normal distributions parameters themselves are sampled from a uniform distribution. We sample $t_i$ uniformly randomly for each event $i$. Thus, the forecast is the value of the true cumulative distribution of the event at $t_i$. We collect ten samples from our synthetic forecasting data at $t=\{0.1T,0.2T,...,0.9T,T\}$.
 
\begin{figure}
    \centering

    \includegraphics[width=0.75\textwidth]{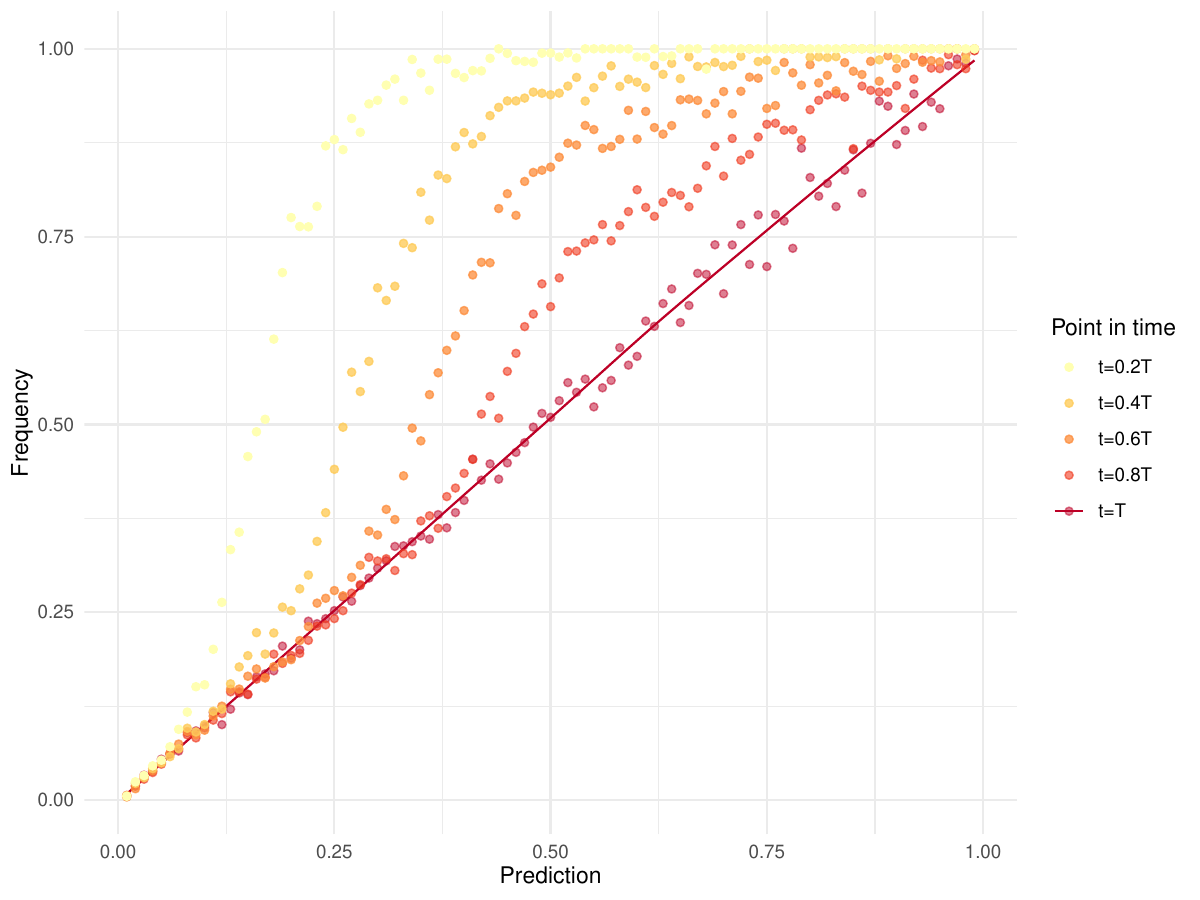}
    \caption{Calibration by time of observation}\label{fig:Calibration_simulation}
\end{figure}

We can see the surprisingly-early bias when we plot the calibration of the forecaster, as measured from different samples (Figure \ref{fig:Calibration_simulation}). 
Although we know that forecasts are perfect, they do not correlate 1-to-1 with real-world frequencies, the points do not fall on the 45-degree diagonal in Figure \ref{fig:Calibration_simulation}, except when $t=T$, when all scheduled resolution dates are in the past.\footnote{The red line, which corresponds to the case $t=T$, is generated using the \texttt{geom\_smooth}-function in R, and is almost exactly the 45-degree diagonal.} The forecasts appear to be systematically too low when $t<T$; positive outcomes in our simulation occur more frequently than predicted, which is what we expected.

\section{Real-world forecasting data}\label{sec:real}

We demonstrate the importance of being aware of the surprisingly-early bias by analyzing real-world forecasting data. We look at Metaculus, a reputation-based massive online prediction-solicitation platform.
The forecasting data on Metaculus is perfectly suited to demonstrate the surprisingly-early bias, 
 as Metaculus features a very diverse set of events---including events whose outcomes have been realized, yet where the scheduled resolution date is after the point of data collection (see Table \ref{tab:Metaculus_example_questions}).

\begin{table}[htbp]
    \centering
    \begin{threeparttable}
    \renewcommand{\arraystretch}{1.3} 
    \caption{Examples of Metaculus questions}
    \label{tab:Metaculus_example_questions}
    \begin{tabularx}{\textwidth}{@{}X X@{}}            
        \toprule
        \multicolumn{2}{c}{\textbf{Event is scheduled to resolve\dots}} \\
        \cmidrule(lr){1-2}
        \textbf{\dots before data collection} & \textbf{\dots after data collection} \\
        \midrule
        Who will win the 2023 Super Bowl? & 
        Will the Taliban capture the Presidential Palace in Kabul by 2026-09-11? \\
        Will NATO Article 5 action be taken before January 1, 2024? & 
        Will the gray wolf be relisted as Threatened or Endangered by the US before 2030? \\
        Will Volodymyr Zelenskyy be named Time Person of the Year in 2022? & 
        Will the United States government either ban TikTok or force a sale before 2025? \\
        \bottomrule
    \end{tabularx}
    \begin{tablenotes}
      \small
      \item \textit{Note.} This table exclusively lists questions on events that did occur, and that therefore have a designated outcome in the forecasting dataset. On the left side of the table we find events whose outcomes (Yes, No) were sure to be verified before data was collected (1st of August 2024). On the right-hand side, we observe events that did occur prior to data collection, although there remained the expectation that they might not have occurred by that point. 
    \end{tablenotes}
    \end{threeparttable}
\end{table}

\begin{figure}
    \centering

    \includegraphics[width=0.95\textwidth]{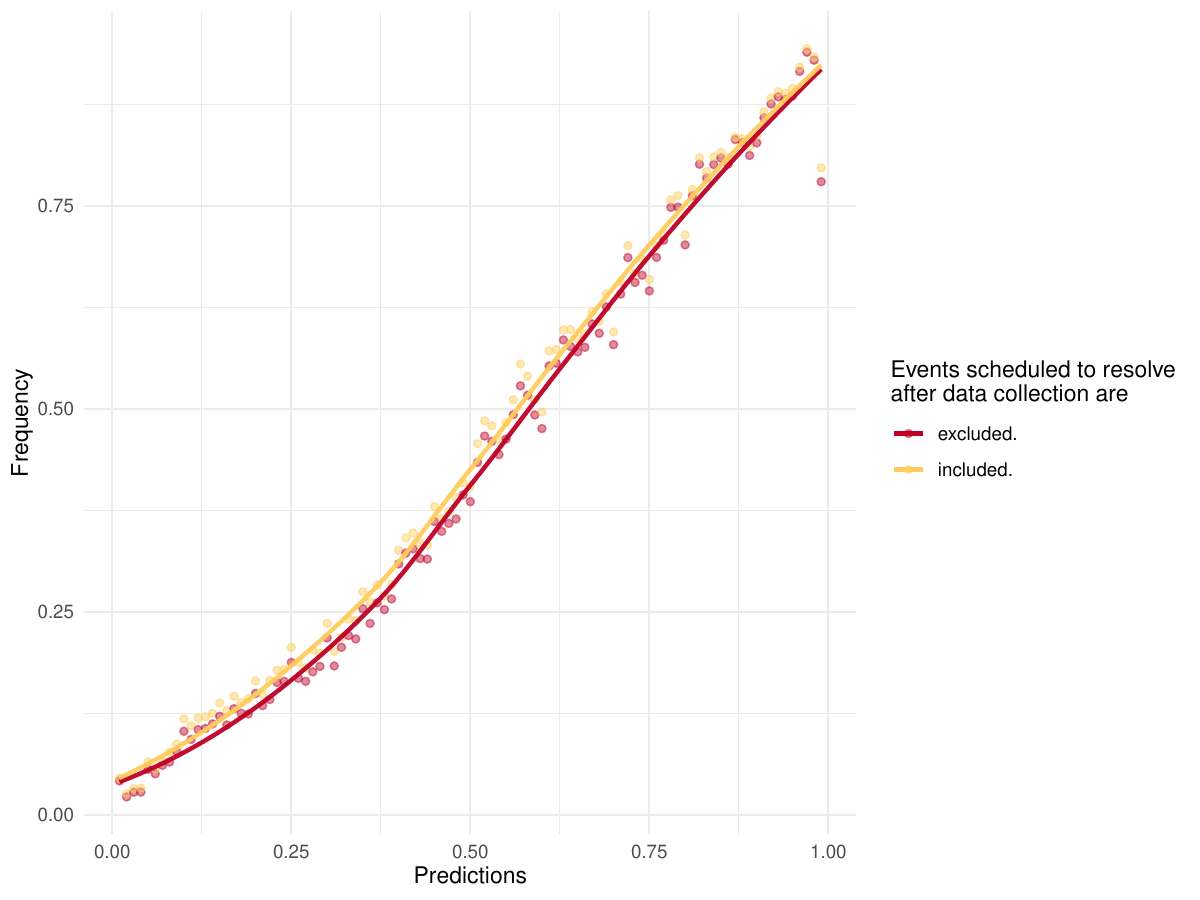}
    \caption{Calibration of Metaculus forecasts}\label{fig:Calibration_metaculus}
    
\end{figure}

We collected 435259 forecasts on 2018 events from 12795 forecasters on Metaculus, the full sample of verified binary probabilistic forecasts available as of August 2024, which is our data collection point. From this dataset we identify 35 events whose scheduled time of resolution $t_i$ is after the data collection point. We plot the calibration of all forecasts with and without these 35 events, see Figure \ref{fig:Calibration_metaculus}, where the horizontal axis contains the different forecasts and the vertical axis refers to the average frequency of the events to which these forecasts corresponded.

Despite the fact that only 1.7\% of the event sample is scheduled to resolve after data collection, we observe a visible difference between the two calibration curves in Figure \ref{fig:Calibration_metaculus}. As in the simulation study from the previous section, the frequency with which positive events occur is---conditional on the forecasts---1.47 percentage points higher when all events are included, which is a significant shift in calibration ($p<0.01$).\footnote{We refer to the supplementary materials for the estimation. It is impossible for this to be lower than 0, yet good to see that it is actually significant. Since the events that are scheduled to resolve after data collection are the ones that have occurred surprisingly early, by definition they must have occurred. Thus, it follows that the observed frequency with which events happen in the biased data is strictly higher than in the \say{clean} data.} Therefore, if we were to do any analysis without excluding events that are scheduled to resolve after data collection, we would run the risk of falsely assessing the accuracy and calibration of predictions.\footnote{Metaculus adjusts for the surprisingly-early bias in their public track record.}

\section{Discussion}

When time-varying events are present in forecasting datasets, observed outcomes can be selected through surprisingly early observation. This can result in misleading evaluations of forecasts and thus---if forecasters (or models) train on feedback---suboptimal forecasting performance. We can draw an analogy to type I and type II errors when evaluating forecasts:

\begin{itemize}
    \item Forecasts may, in reality, be well calibrated, yet be incorrectly judged as miscalibrated.
    \item Conversely, forecasts that are systematically miscalibrated might have this flaw at least partially concealed.
\end{itemize}

In order to eliminate the surprisingly-early bias in forecasting data, we need to make sure that all forecasts have been scheduled to resolve in the past, such that their presence in the dataset does not depend on the time of observation. 
This may very well mean that we need to discard data for existing and verified forecasts, as demonstrated in section \ref{sec:real}, or that we need to wait to evaluate forecasts, as with the study of \textcite{field_forecasting_2008} who report the cumulative probability that earthquakes occur in California over the next 30 years. 
In this case we may not want to wait with the evaluation until all events are verified, for this may take a long time and decisions need to be made now, so we may try to impute the likely density forecast by making assumptions, e.g. that the yearly risk of earthquakes is constant across time.

The superior approach is to collect density forecasts across time. Had \textcite{field_forecasting_2008} instead shared the probability density function---specifying the probability of earthquakes in each individual year--- we would always be able to extract the forecasts for earthquakes up until the data collection point.

\printbibliography

\end{document}